\documentclass[a4paper,12pt, twoside]{article}
\usepackage{amsmath,float,bm}
\usepackage{hyperref}
\usepackage{authblk,graphicx}

\usepackage[total={6.5in,8.75in},
top=1.2in, left=0.9in, includefoot]{geometry}

\begin{document}

\title{\textbf{Chaotic trajectories in complex Bohmian systems}}

\author{A.C. Tzemos \footnote{Corresponding Author: atzemos@academyofathens.gr} and G. Contopoulos}
\affil{Research Center for Astronomy and 
Applied Mathematics of the Academy of 
Athens - Soranou Efessiou 4, GR-11527 Athens, Greece}

\maketitle

\begin{abstract}
We consider the Bohmian trajectories in a 2-d quantum harmonic oscillator with non commensurable frequencies whose wavefunction is of the form $\Psi=a\Psi_{m_1,n_1}(x,y)+b\Psi_{m_2,n_2}(x,y)+c\Psi_{m_3,n_3}(x,y)$.  We first find the  trajectories of the nodal  points for different combinations of the quantum numbers $m,n$. Then we study, in detail, a case with relatively large quantum numbers and  two equal $m's$. We find 
(1) fixed nodes independent of time and (2) moving  nodes which from time to time collide with the fixed nodes and at particular times they go to infinity. 

Finally, 
we study the trajectories of quantum particles close to the nodal points and observe, for the first time, how chaos is generated in a complex system with multiple nodes scattered on the configuration space. 

\end{abstract}

\section{Introduction}\label{sec1}
Bohmian Quantum Mechanics (BQM) \cite{Bohm,BohmII,holland1995quantum} is one of the main interpretations of Quantum Mechanics, where the quantum particles follow trajectories according to the Bohmian equations of motion:
\begin{align}
m\frac{dr}{dt}=\hbar\Im\left(\frac{\nabla\Psi}{\Psi}\right),
\end{align}
where $\Psi$ is the wavefunction, i.e. the solution of Schr\"{o}dinger's equation
\begin{equation}
\frac{-\hbar^2}{2m}\nabla^2\Psi+V\Psi=i\hbar\frac{\partial \Psi}{\partial t}.
\end{equation}

As it is well known, the trajectories of  quantum particles in BQM are either ordered or chaotic. Chaos is established when a trajectory approaches the neighbourhood of a nodal point, where the wavefunction $\Psi=0$ \cite{parmenter1995deterministic, iacomelli1996regular,frisk1997properties,konkel1998regular, wu1999quantum,
falsaperla2003motion, wisniacki2003dynamics, wisniacki2005motion,efthymiopoulos2006chaos,
wisniacki2007vortex, cesa2016chaotic}.

In fact,  in \cite{efthymiopoulos2007nodal, efth2009} it was shown that, close to every nodal point of a arbitrary wavefunction, there is at least one hyperbolic fixed point,  in the frame of reference of the node, the so called `X-point'. The X-point has two opposite unstable and two opposite stable eigendirections and together with the nodal point form the so called 'nodal point-X-point complex' (NPXPC), a distinctive form of the Bohmian flow in the neighbourhood of the nodal point. Particles approaching the X-point along trajectories close to a stable direction are deviated by the X-point along the two opposite unstable directions. The result of many such scattering events is the emergence of chaos in the Bohmian trajectories.

Some Bohmian trajectories are trapped in the region of the moving nodal point and form vortices around it, but later, when the nodal point acquires large velocity, they escape far away from its neighbourhood. On the other hand, trajectories beyond the X-point do not form any loops around the nodal point.

\begin{figure}
\centering
\includegraphics[width=0.7\textwidth]{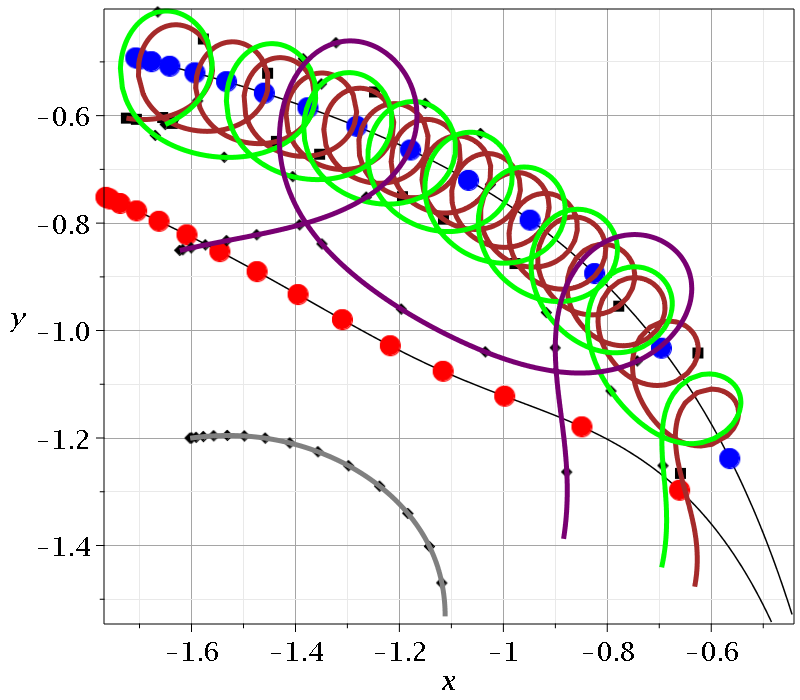}
\caption{Four Bohmian trajectories in the case of the wavefunction $\Psi=a\Psi_{0,0}+b\Psi_{1,0}+c\Psi_{1,1}$, with $a=b=1$ and $c=\sqrt{2}/2$, which has    a single nodal point, together with the trajectories of the nodal point (connected blue dots) and of the X-point (connected red dots) in the inertial frame $(x,y)$ for a time interval $t\in[0.01,1.5]$. The initial conditions of the trajectories are $\{x(0.01)=-1.7252, y(0.01)=-0.6045\}$ (brown),
$\{x(0.01)=-1.6504, y(0.01)=-0.6166\}$ (green), $\{x(0.01)= -1.6225, y(0.01)=-0.8510\}$ (purple), and $ \{x(0.01)=-1.6026, y(0.01)=-1.2004\}$ (grey). The trajectories that approach the nodal point form loops around it for some time (the closer the approach the larger the time of the spiral motion. Trajectories that stay away from the X-point (grey) do not form any loops around the nodal point.}\label{fig1}
\end{figure}

The NPXPC mechanism is generic, namely it is applicable to any quantum state and has been extensively tested  in 2-d quantum harmonic oscillators with incommensurable frequencies (see also \cite{borondo2009dynamical}).  However, in most cases studied so far, the wavefunctions had one, two or three nodal points, i.e. they were quite simple. Working with small quantum numbers  was necessary  for the analytical calculation of the positions of the nodal points in the configuration space, as a function of time $t$.

The theory was further extended
in cases of 3-d Bohmian systems where we found the  `3-d structures of NPXPCs' \cite{tzemos2016origin, contopoulos2017partial, tzemos2018integrals, tzemos2018origin}.

The first work towards more realistic cases of both theoretical and technological significance was made in \cite{tzemos2019bohmian,tzemos2020chaos,tzemos2020ergodicity,tzemos2021role}, where we studied the Bohmian trajectories of two entangled qubits made of coherent states of the QHO. There we found that there is an important interplay between order, chaos and entanglement, which leads to a better understanding of the dynamical establishment of Born's rule $P=\|\Psi\|^2$, in cases where initially $P_0\neq\|\Psi_0\|^2$ \cite{valentini1991signalI, valentini1991signalII, valentini2005dynamical}.  That system had infinitely many NPXPCS on a straight line moving and rotating in the configuration space. The remarkable property of that system was that we could find analytically the positions of every nodal point as a function of time $t$.

However, in a generic Bohmian quantum system, one expects to have multiple NPXCs  distributed inside the support of the wavefunction (the region where $P=\|\Psi\|^2$ is not negligible).

Thus in the present paper we work with a wavefunction of  more complex systems, whose nodal points are scattered on the $x,y$ plane.  We consider various types of wavefunctions that have many scattered nodal points. Then we study in detail a typical example which has  both moving and fixed nodal points. We find that the moving nodes collide from time to time with the fixed nodes, something that changes drastically the behaviour  of the Bohmian trajectories of particles passing close to them. Thus we observe the evolution of the NPXPCs in a case which has a fairly realistic complexity.

Finally we make a short comment on the relation between the position of the X-points and the quantum potential and the total potential as well. We find numerically that the X-points of our model are close to the local maxima of the quantum potential and of the total potential.

The structure of the paper is the following: In section 2 we present the model and  we make a classification of the nodal points according to different combinations of the quantum numbers. In section 3 we study a typical example where we find detailed results on the evolution of the NPXPCs.  In section 4 we examine the behaviour of the  trajectories of quantum particles 
in our example and show how important is the interaction between  the two types of nodal points for the form of Bohmian trajectories. In section 5 we present a short numerical result for the form of both the quantum potential and the total potential close to the nodal points and  the X-points and far from them. In section 6 we consider the cases of commensurable frequencies. In section 7 we make a summary of our results and draw our conclusions. Finally, in the appendix we discuss the way of finding the nodal points in complex wavefunctions.




\section{Trajectories of nodal points}\label{sec2}
We consider the simple case of a 2-d harmonic oscillator
\begin{align}
V=\frac{1}{2}(\omega_1^2x^2+\omega_2^2y^2),
\end{align}
where $\omega_1/\omega_2$ is an irrational number. The basic solutions in the 1-d case are  the well known wavefunctions of the quantum harmonic oscillator,
\begin{equation}
\Psi_{m}=e^{-i\phi}e^{-\frac{\omega_1x^2}{2}}H({x^m}),
\end{equation}
where  $\phi=(\frac{1}{2}+m)\omega_1t$ and $H(x^m)$ is a Hermite polynomial of degree $m$.
At the nodal points we have:
\begin{equation}
\Psi=\Psi_{Real}+i\Psi_{Imaginary}=0
\end{equation}

We consider wavefunctions which are linear combinations of some basic solutions written in the form:
\begin{equation}
\Psi=a\Psi_{m_1,n_1}+b\Psi_{m_2,n_2}+c\Psi_{m_3,n_3}
\end{equation}
The partial solutions $\Psi_{m,n}$ are of the form\footnote{In fact we have \begin{equation}
\Psi_{m,n}=\Psi_m(x)\cdot\Psi_n(y)
\end{equation}
where 
\begin{align}
\Psi_l(x)=\frac{1}{\sqrt{2^ll!}}\left(\frac{m_x\omega_x}{\pi\hbar}\right)^{\frac{1}{4}}e^{-\frac{m_x\omega_x x^2}{2\hbar}}H_l\left(\sqrt{\frac{m_x\omega_ x}{\hbar}}x\right),l=0, 1, 2,\dots,
\end{align} and similarly for $y$.}
\begin{equation}
\Psi_{m,n}=e^{-i\phi_{m,n}}e^{-\frac{\omega_1x^2+\omega_2y^2}{2}}H({x^m})H(y^n),
\end{equation}
where $H(x^m)$ are Hermite polynomials  of degree $m$ in $x$ and $n$ in $y$ and the angle $\phi$ is equal to 
\begin{equation}
\phi_{m,n}=\left[\left(\frac{1}{2}+m\right)\omega_1+\left(\frac{1}{2}+n\right)\omega_2\right]t.
\end{equation}
The equations that give the nodal points are
\begin{eqnarray}
\nonumber &aH(x^{m_1})H(y^{n_1})\cos(\phi_1)+bH(x^{m_2})H(y^{n_2})\cos(\phi_2)\\&+cH(x^{m_3})H(y^{n_3})\cos(\phi_3)=0\label{s1}\\
\nonumber&aH(x^{m_1})H(y^{n_1})\sin(\phi_1)+bH(x^{m_2})H(y^{n_2})\sin(\phi_2)\\&+cH(x^{m_3})H(y^{n_3})\sin(\phi_3)=0\label{s2}
\end{eqnarray}
The Hermite polynomials $H(x^m)$ or $H(y^n)$, are of degree $m$ or $n$ and contain terms of degrees $m, m-2, m-4...$ or $n, n-2, n-4,\dots$.

\subsection{Special Cases}

\begin{enumerate}
\item If two $m's$ are equal e.g. $ m_1=m_2$ we have 
\begin{eqnarray}
&H(x^{m_1})[aH(y^{n_1})\cos(\phi_1)+bH(y^{n_2}\cos(\phi_2)]+cH(x^{m_3})H(y^{n_3})\cos(\phi_3)=0\label{q1}\\&
H(x^{m_1})[aH(y^{n_1})\sin(\phi_1)+bH(y^{n_2})\sin(\phi_2)]+cH(x^{m_3})H(y^{n_3})\sin(\phi_3)=0\label{q2}
\end{eqnarray}
A set of solutions of these equations is $H(x^{m_1})=H(y^{n_3})=0$. These solutions are independent of the angles, therefore they are fixed on the plane $(x,y)$. Beyond these solutions we find further solutions if  we multiply Eq.\eqref{q1} by $[aH(y^{n_1})\sin(\phi_1)+bH(y^{n_2})\sin(\phi_2)]$ and 
Eq.\eqref{q2} by $[aH(y^{n_1})\cos(\phi_1)+bH(y^{n_2})\cos(\phi_2)]$ and  subtract the two equations. We find
\begin{equation}
H(x^{m_3})\left[{aH(y^{n_1})\sin(\phi_1-\phi_3)+bH(y^{n_2})\sin(\phi_2-\phi_3)}\right]=0\label{q3}
\end{equation}
The general solution of Eq.\eqref{q3} gives $y$ as a trigonometric function of the time. Then we can use the solution $y=y(
\phi_1,\phi_2,\phi_3)$ in Eq.\eqref{q1} or \eqref{q2} and find $x$ as a trigonometric function of $t$.
Similar results are found in the case where two $n's$ are equal. 

Therefore if two $m's$ or $n's$ are equal we find the solutions for $x,y$ which are functions of the angles, that are proportional to $t$. These solutions can be given analytically if $m$ and $n$ are at most 4. Otherwise the solutions can be found only numerically.

\item If three $m's$ are equal, then the function $\Psi$ is of the form
\begin{equation}
\Psi=e^{\frac{-\omega_1x^2+\omega_2y^2}{2}}H(x^m)\left[a\Psi(y^{n_1})+b\Psi(y^{n_2})+c\Psi(y^{n_3})\right]
\end{equation}
and the nodal points are at fixed values of $x$ which are given by the equation $H(x^m)=0$, while the solutions $y=y(t)$ depend on the time $t$ through the angles $\phi_1,\phi_2,\phi_3$. Therefore the problem is 1-dimensional and there is no chaos.

\item In the most general case, all the $m's$ and $n's$ are different. In such cases it is not possible, in general, to find analytical solutions for the nodal points.
However, we can find analytical solutions if we have relatively small (but not equal) $m's$ and $n's$. E.g. if $m_1=0, m_2=1$ and $m_3=2$ we can eliminate the third term by multiplying Eq.\eqref{s1} by $\sin(\phi_3)$ and Eq.\eqref{s2} by $\cos(\phi_3)$ and subtracting the two terms. Then we find 
\begin{equation}
H(x^0)H(y^{n_1})\sin(\phi_3-\phi_1)+H(x^1)H(y^{n_2})\sin(\phi_3-\phi_2)=0,
\end{equation}
where $H(x^0)=1, H(x^1)=2x$ and thus 
\begin{equation}
x=-\frac{H(y^{n_2})\sin(\phi_3-\phi_2)}{H(y^{n_1})\sin(\phi_3-\phi_1)}.\label{x3}
\end{equation}
Introducing this value in Eq. \eqref{s1} or \eqref{s2} we find an equation that contains only $y$ and the angles, namely we find $y(\phi_1, \phi_2, \phi_3)$. Its solution gives the values of $y$ as functions of time $t$ and then Eq.\eqref{x3} gives the corresponding values of $x$.
%

However the applicability of this method is restricted on rather small values of the quantum numbers, because for high order Hermite polynomials, we cannot isolate, in general, $y(\phi_1,\phi_2,\phi_3)$.

\end{enumerate}

\section{A typical example with fixed and moving nodal points}\label{sec3}
We consider the case of the wavefunction
\begin{equation}
\Psi=a\Psi_{3,3}+b\Psi_{3,4}+c\Psi_{4,5}
\end{equation}
This wavefunction\footnote{This is an example of a non stationary state with stationary nodes and chaos. This answers a question raised by Cesa, Martin and Struyve \cite{cesa2016chaotic}, who could not find a system of this type and wondered if such a system would generate chaos.} has two equal $m's$ ($m_1=m_2=3$). In the 
numerical calculations we take $a=b=1,c=\sqrt{2}/2$ and $\omega_1=1, \omega_2=\sqrt{2}/2$.
The nodal points satisfy the equations
\begin{eqnarray}
&H(x^3)\left[aH(y^3)\cos(\phi_1)+bH(y^4)\cos(\phi_2)\right]+cH(x^4)H(y^5)\cos(\phi_3)=0\label{l1}\\&
H(x^3)\left[aH(y^3)\sin(\phi_1)+bH(y^4)\sin(\phi_2)\right]+cH(x^4)H(y^5)\sin(\phi_3)=0,\label{l2}
\end{eqnarray}
where
$\phi_1=\frac{7}{2}(1+\omega_2)t,\phi_2=\frac{1}{2}(7+9\omega_2)t,\phi_3=\frac{1}{2}(9+11\omega_2)t$.
One type of solutions is 
$H(x^3)=-\frac{2}{3}x^3+x=0$  and 
$H(y^5)=-(\frac{4}{15}\omega_2y^4-\frac{1}{3}\omega_yy^2+1)\sqrt{\omega_2y}=0$.
Thus $x=0$, $x=\pm\sqrt{3}/2$ and $y=0$, $y=\pm\frac{1}{\sqrt{2\omega_2}}\sqrt{5\pm\sqrt{10}}$. These solutions are time independent (they are fixed in the $(x,y)$ plane). There are 5 sets of triplets with the same $y$ and $x=0, \pm\sqrt{3}/2$, namely  15 fixed (time independent) solutions fixed in the configuration space $(x,y)$

Besides these solutions we find also the following equations by eliminating the first two terms of Eqs.\eqref{l1}-\eqref{l2} as in the case  1. above
\begin{equation}
H(x^4)H(y^4)[aH(y^3)\sin(\phi_1-\phi_3)+bH(y^4)\sin(\phi_2-\phi_3)]=0\label{l3}
\end{equation}
or
\begin{equation}
H(y^3)\sin[(1+2\omega_2)t]+H(y^4)\sin[(1+\omega_2)t]=0.
\end{equation}
This equation gives four new time dependent solutions for $y$. Introducing these solutions in Eq.\eqref{l1} or \eqref{l2} we find an equation of degree 4 in $x$ with four time dependent solutions. 
The number of theses solutions is 16 but as we want only real solutions we may have a smaller number of nodes. We note, that in the present paper all solutions are real.

\begin{figure}
\centering
\includegraphics[width=0.7\textwidth]{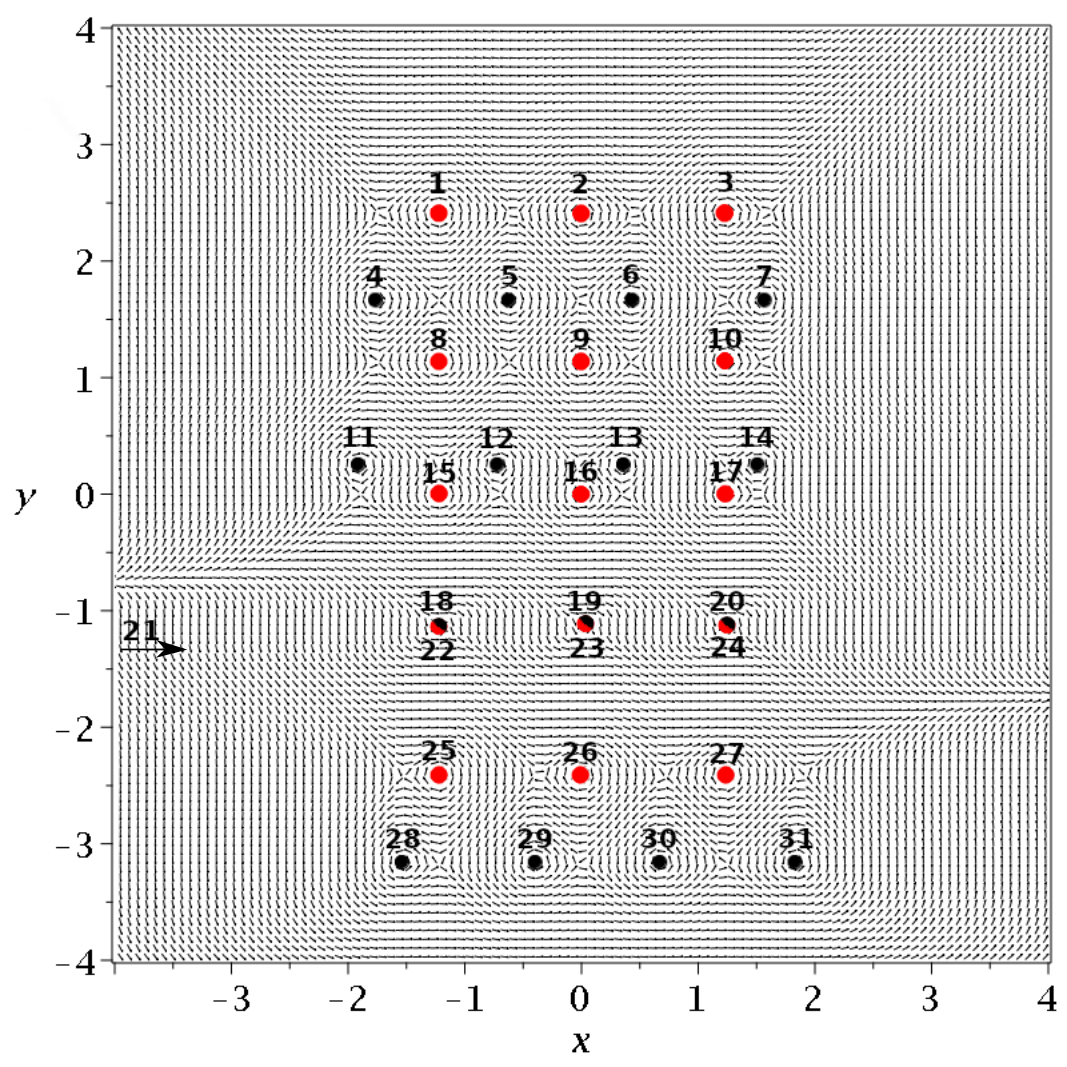}
\caption{The set of nodes for $t=0.1$ numbered sequentially. The fixed nodes are red and the moving nodes are black. The node 21 is out of this frame but it comes closer as $t$ increases.}\label{fig2}
\end{figure}

The total number of solutions is $16+15=31$. In Fig.~\ref{fig2} we show the $30$ nodal points at time $t=0.1$,  $15$ fixed and $15$ moving. The missing solution (number 21) is on the left and  outside this figure.

\begin{figure}
\centering
\includegraphics[width=0.7\textwidth]{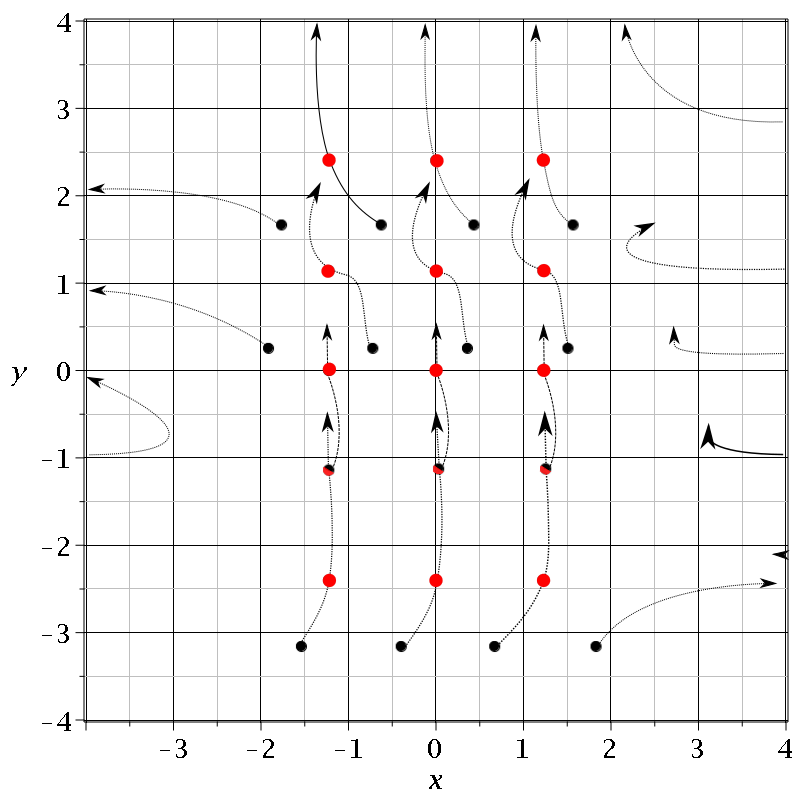}
\caption{The nodal trajectories (with arrows) in the time interval $t\in[0.1,2.5]$}\label{fig3}
\end{figure}

The time dependent solutions appear in sets of quadruplets with the same $y$ and they move upwards (Fig.~\ref{fig3}). When $\sin[(1+\omega_2)t]=0$ the set $4, 5, 6, 7$ goes to $y=\infty$. This happens when $t=k\pi/(1+\omega_2)=1.84k$. The other three sets have $y$ equal to the roots of $H(y^3)=0$, i.e. $y=0$ and $y=\pm\sqrt{\frac{3}{2\omega_2}}=\pm 1.4565$. Later the set $4, 5, 6, 7$ reappears from $y=-\infty$ upwards.

\begin{figure}
\centering
\includegraphics[width=0.7\textwidth]{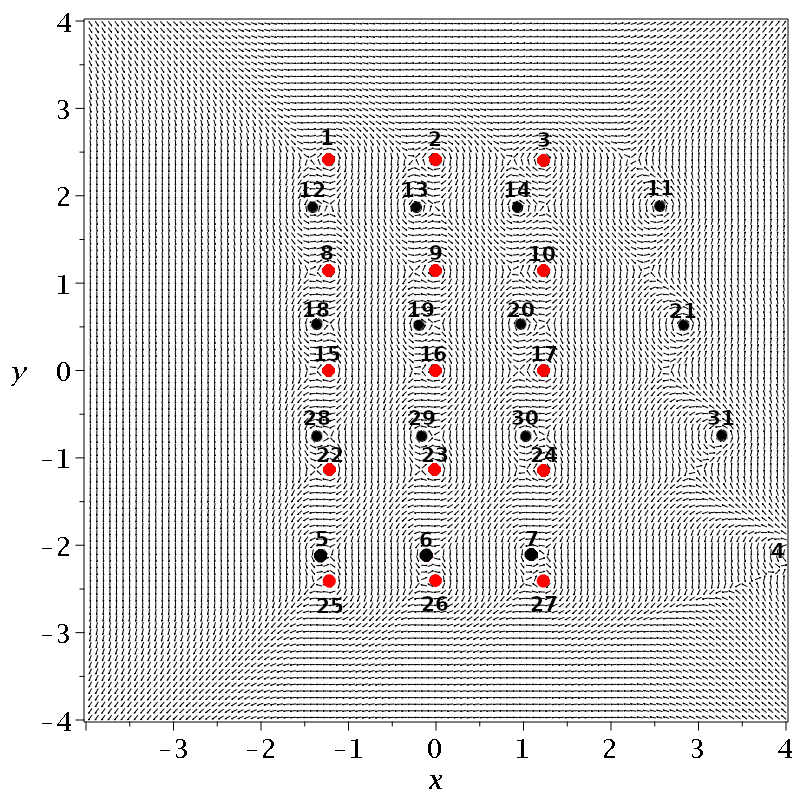}
\caption{The positions of the nodal points for $t=2.5$. While the fixed nodal points do not change, the moving nodal points have changed completely in comparison with their positions in Fig.~\ref{fig2} ($t=0.1$).}\label{fig4}
\end{figure}

At $t=\frac{\pi}{1+\omega_2}$ and $y=0$, the equations \eqref{l1} and \eqref{l2} have $H(y^6)=0$ and $H(x^3)=0$. Therefore the 3 moving nodes $18, 19, 20$ coincide with the  fixed nodes $15, 16 17$, while the fourth moving node (No 21) comes from $x=-\infty$ to a maximum $x$ about $x=-3$ and  moves again  to $x=-\infty$. Later this node reappears coming from $x=\infty$. The positions of the nodes at $t=2.5$ are shown in Fig.~\ref{fig4}

The nodes $4, 11, 31$ go to infinity in the same way when the corresponding moving sets $(5, 6, 7), (12, 13, 14), (28, 29, 30)$ collide with some sets of 3 fixed nodes.

Collisions of nodes occur also at some times besides $t=k\pi/(1+\omega_2)$. E.g. when $t=1.47551$ one moving solution of Eq.\eqref{l3} is $y=2.4024$ and this value gives $H(y^5)=0$. Then Eq.\eqref{l1} or \eqref{l1} gives for $x$ the 3 roots of $H(x^3)=0$. And in fact, the solutions for the nodal points $5, 6,$ and $7$ are $-\sqrt{3}/2, 0, \sqrt{3/2}$, i.e. these nodal points collide with the fixed nodal points $1, 2, 3$. Finally, the positions of the node 4 goes to $x=-\infty$ as $t$ approaches this particular value, and beyond that value it appears again with positive $x$'s, coming from $x=\infty$.

\section{Trajectories of quantum particles}

\begin{figure}
\centering
\includegraphics[width=0.7\textwidth]{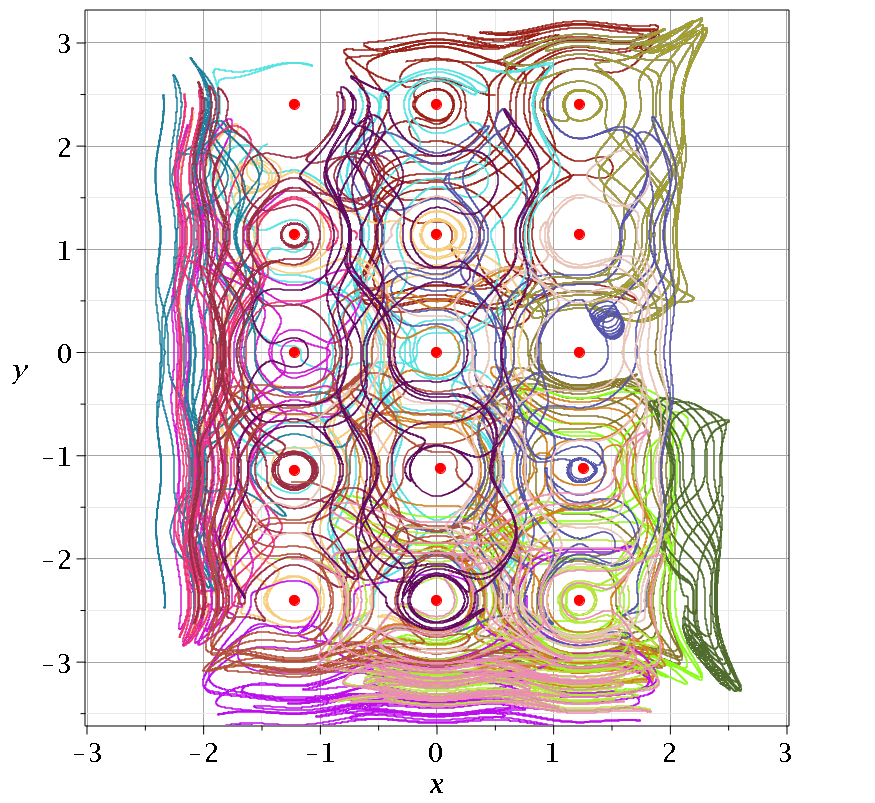}
\caption{Various trajectories of Bohmian particles in the time interval $t=[0.1,60]$. The red dots represent fixed nodal points. All the trajectories inside the grid of the nodal points are chaotic in the long run.}\label{fig5}
\end{figure}

The trajectories of most particles that populate the central region of the configuration space $(x,y)$ are quite irregular (chaotic). This is seen in Fig.~\ref{fig5} which contains many trajectories for a time interval from $t=0.1$ up to $t=60$. Only trajectories remaining outside the central region $[-2, 2]\times[-3, 3]$ may be ordered.

\begin{figure}
\centering
\includegraphics[width=0.8\textwidth]{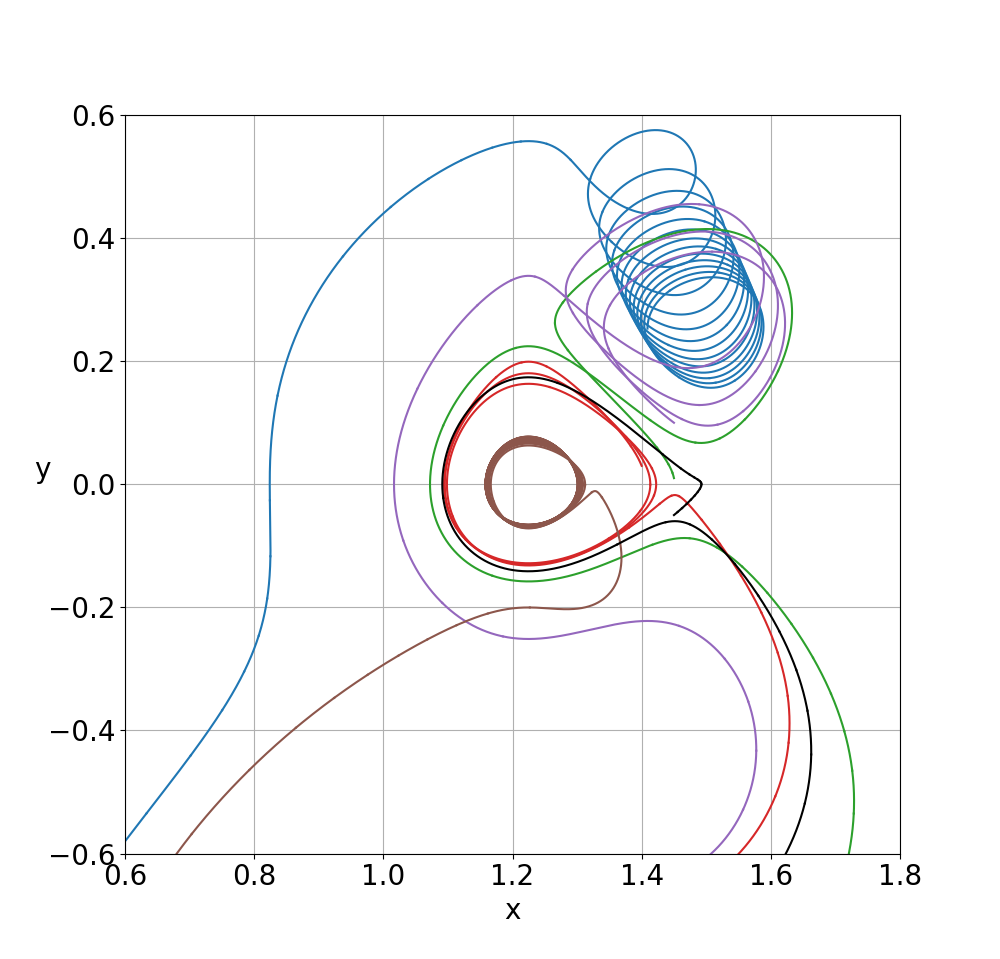}
\caption{Trajectories of quantum particles starting near the nodal points 14 and 17 for $t\in[0.1,2.5]$ (Initial conditions: \{x(0.1)=1.409, y(0.1)=0.253 (blue), x(0.1)=1.45, y(0.1)=0.01 (green),  x(0)=1.40, y(0)=0.03 (red), x(0.1)=1.45, y(0.1)=0.1 (purple), x(0.1)=1.3, y(0.1)=0.01 (brown) and x(0.1)=1.45, y(0.1)=-0.05 (black)\}).}\label{fig6n}
\end{figure}

 However, the greatest interest is in the the trajectories that are trapped for some time around central nodal points. 
An example of trajectories close to the moving node 14 and the fixed node is given in Fig.~\ref{fig6n}. The trajectories that start close to the moving node form loops around it for some time. But later they deviate to large distances from it. The initial time of the trajectories is $t=0.1$. The time of trapping around the moving node is larger when the initial distance of the trajectory from the nodal point is smaller.

\begin{figure}
\centering
\includegraphics[width=0.8\textwidth]{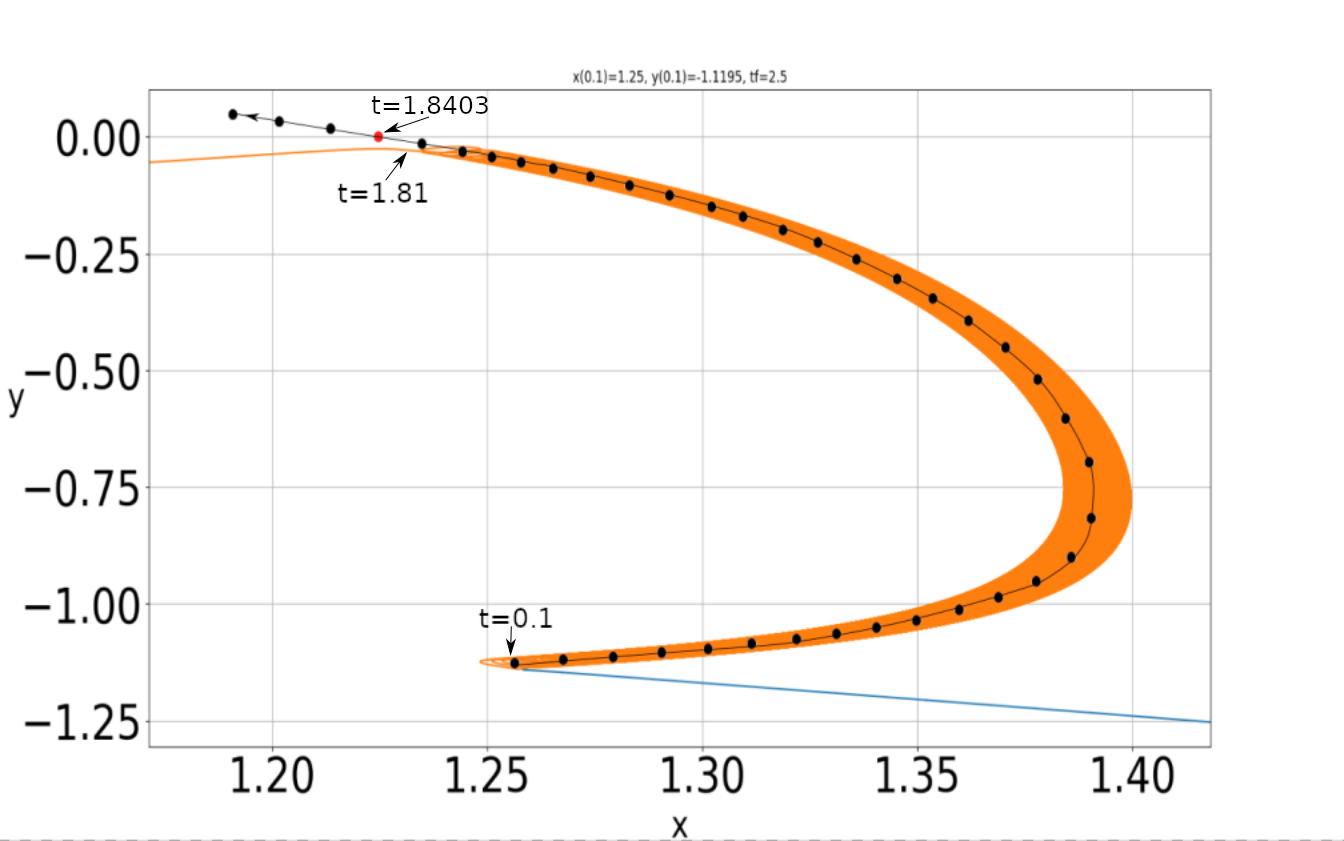}
\includegraphics[width=0.85\textwidth]{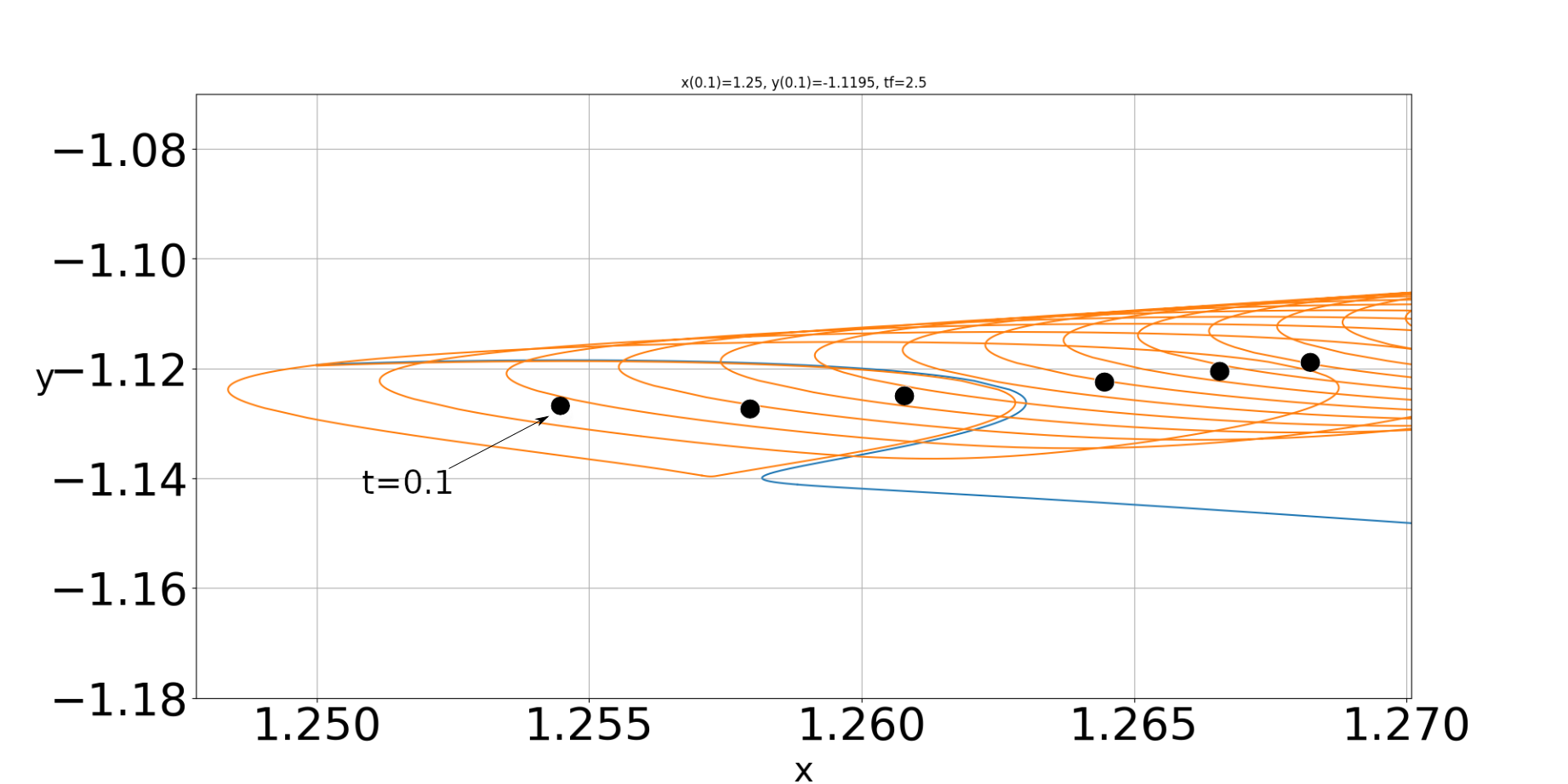}
\caption{a) The trajectory of the moving nodal point No20 (connected black dots) for a time interval (0.1-1.85). This node is initially at $x_N(0.1)=1.2544, y_N(0.1)=-1.1264$ and reaches the fixed node 17  ($x=1.2245, y=0$) at $t=1.8403$. Two trajectories of particles close to it are with initial conditions at $t=0.1$ $x(0.1)=1.25, y(0.1)=-1.1193$ (orange) and $x(0.1)=1.25, y(0.1)=-1.1193$ (blue). The orange trajectory forms a large number of nearby loops. b) A zoom of the initial arcs of the trajectories of the nodal point 20 and of the trajectories of particles. The two trajectories of particles are close until they approach an X-point 
and then they deviate into opposite directions.}\label{fig7}
\end{figure}

The blue curve starts closer to the node 14 and forms 13 loops before escaping. The purple trajectory starts further away from the node 14 and makes 3 loops around the node 17. Then it makes a large loop around the node 14 and escapes downwards. The green curve starts even further away from the node 14 and makes only one loop around it and a loop around the node 17 before escaping downwards.
The black curve makes only one loop around the node 17 and escapes downwards. Finally the red and the brown curves make a number of loops around the node 17 before escaping downwards (the brown curve is closer to the node 17 and makes more loops around it). 
As time goes beyond $t=1.25$, the moving node 14 goes far from the fixed node 17, but another moving node (node 20) approaches the fixed node 17, and reaches it at $t=1.8403$ (Fig.~\ref{fig7}a).

\begin{figure}
\centering
\includegraphics[width=0.7\textwidth]{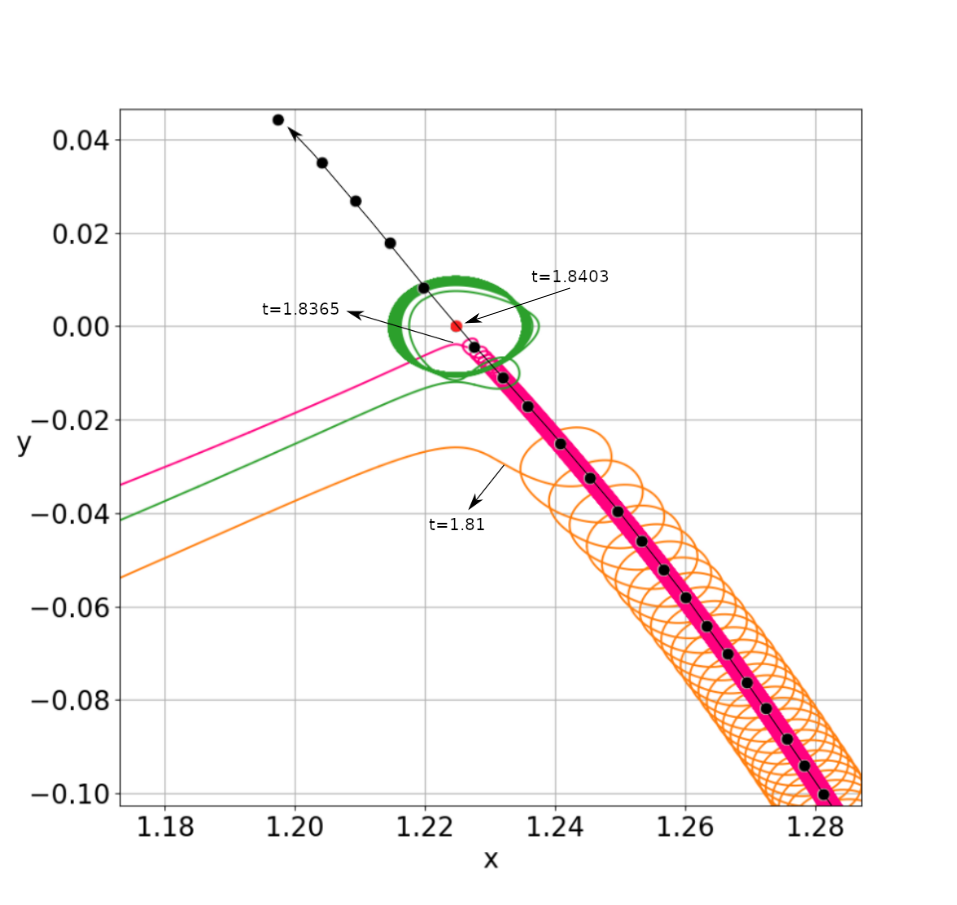}
\caption{ 
Two trajectories starting close to the nodal point 20. The first one is the orange curve of Fig.\ref{fig7}, while the second one (magenta) is much closer to the nodal point (with  $x(0.1)=1.254, y(0.1)=-1.121$ and forms loops up to $t=1.8365$. The trajectory of the nodal point 20 is again given by connected black dots. Finally the green trajectory which starts at $t=0.1$ close to the nodal point 14 ($x(0.1)=1.228, y(0.1)=0.01$) makes loops around the node up to $t\simeq 1.83$.
}\label{fig8}
\end{figure}

Trajectories that start close to the node 20 form a large number of loops around it. E.g. the orange trajectory, starting at $x(0.1)=1.25, y(0.1)=-1.1195$ escapes from the neighborhood of the node 20 at the time $t\simeq 1.81$ (Fig.~\ref{fig7}a). Trajectories starting a little further away from the node (blue curve starting at $x=1.25, y=-1.1193$) do not make any loops around the node 20 but escape to the right. In Fig.~\ref{fig7}b we see that the  orange and blue curves are very close for some time, but when they approach an X-point the orange curve goes to the left and later forms loops, while the blue curve turns right to large distances. This is  a typical example of chaos generation close to an X-point.
When the moving node approaches the fixed node, this trajectory escapes. This happens at $t\simeq1.81$ (Fig.~\ref{fig7}a and in greater detail in Fig.~\ref{fig8}).

Trajectories that start closer to the node 20 at $t=0.1$, remain close to it for a little longer time. In Fig.~\ref{fig8} we draw the arc (magenta) of a trajectory starting at $x(0.1)=1.254, y(0.1)=-1.121$. 
This trajectory follows the node 20 up to $t=1.836$ and then it escapes. Close to the fixed node there is a trajectory (green) starting at $x(0.1)=1.228, y(0.1)=0.01$ that  forms a ring around the fixed node up to about $t=1.83$ and then escapes. Of course, it is not possible for a trajectory to remain close to the fixed node when a moving node comes very close to it.

\begin{figure}
\centering
\includegraphics[width=0.7\textwidth]{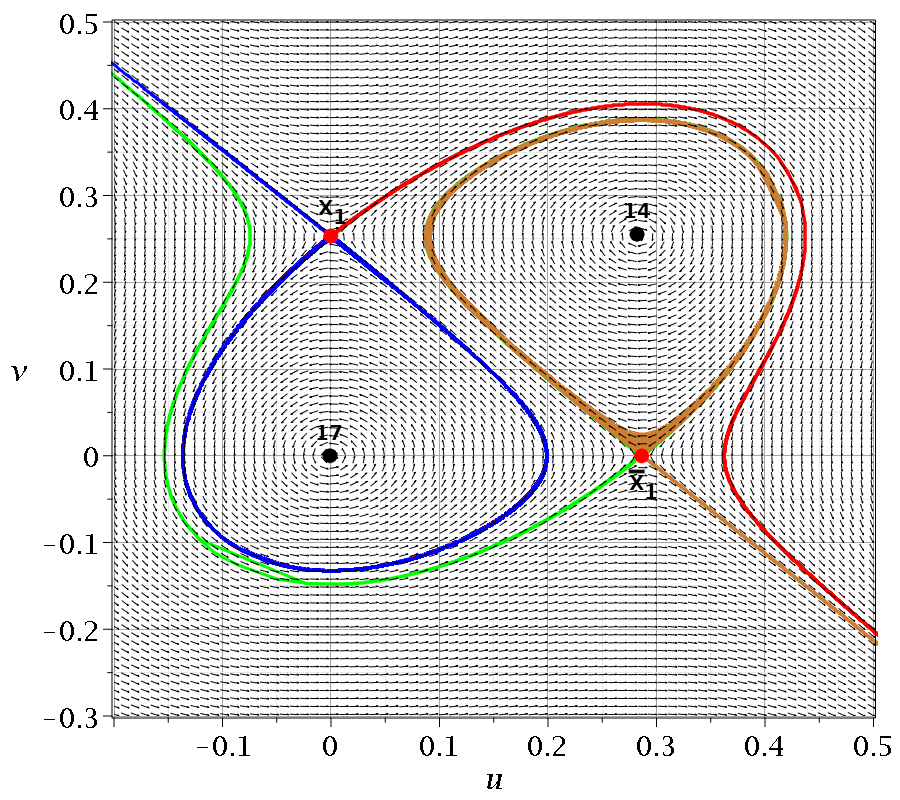}
\caption{ The flow on the plane $(x,y)$ close to the nodes 17 (fixed) and 14 (moving) at the time t=0.1. There are two unstable stationary points ${X}_1$ and $\bar{X}_1$ from which start stable and unstable asymptotic curves.}\label{flow}
\end{figure}

\begin{figure}
\centering
\includegraphics[width=0.7\textwidth]{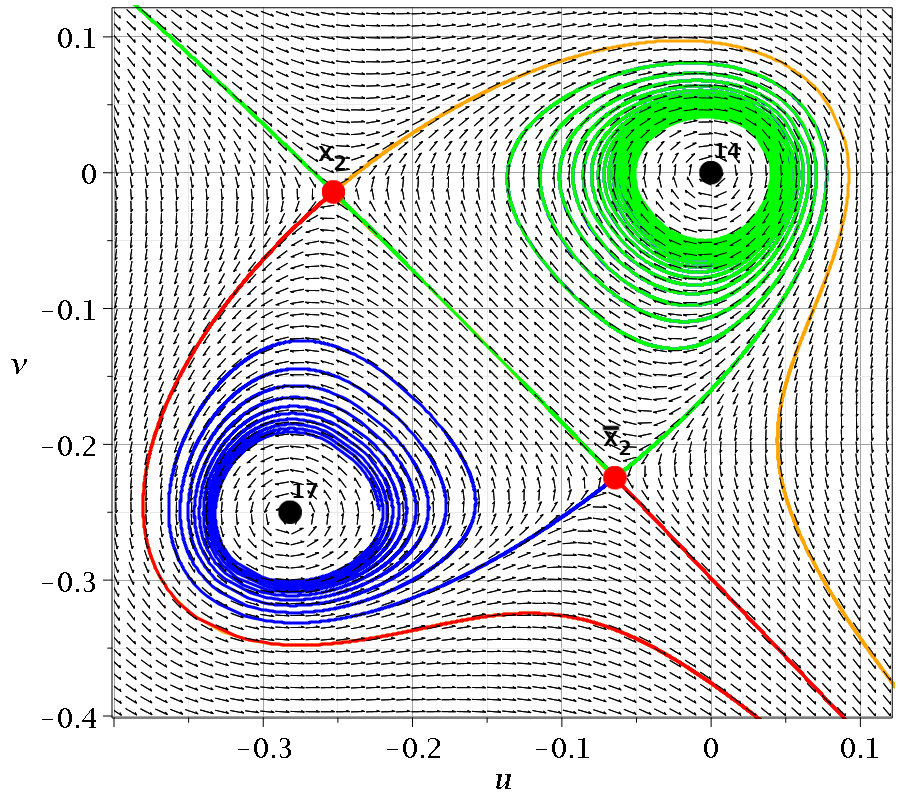}
\caption{ As in Fig.~\ref{fig9} but in a coordinate system $(u,v)$ centred at the moving nodal point. In this system there are two X-points ${X}_2$,$\bar{X}_2$, that are close to ${X}_1$ and $\bar{X}_1$, but different from them.}\label{fig10}
\end{figure}

\begin{figure}
\centering
\includegraphics[width=0.45\textwidth]{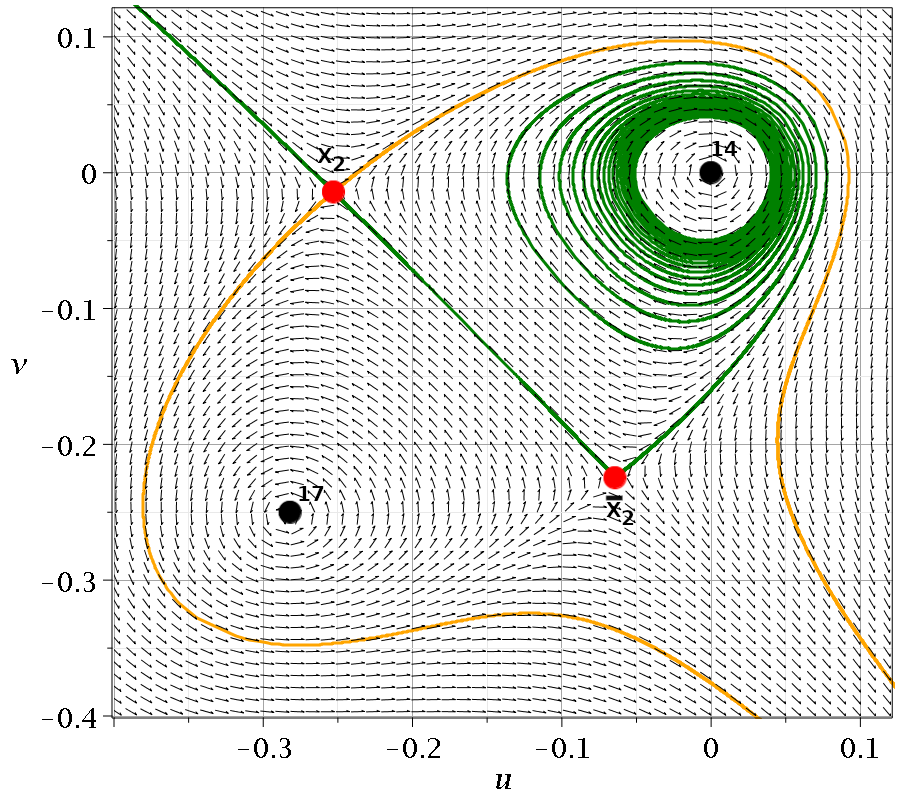}
\includegraphics[width=0.45\textwidth]{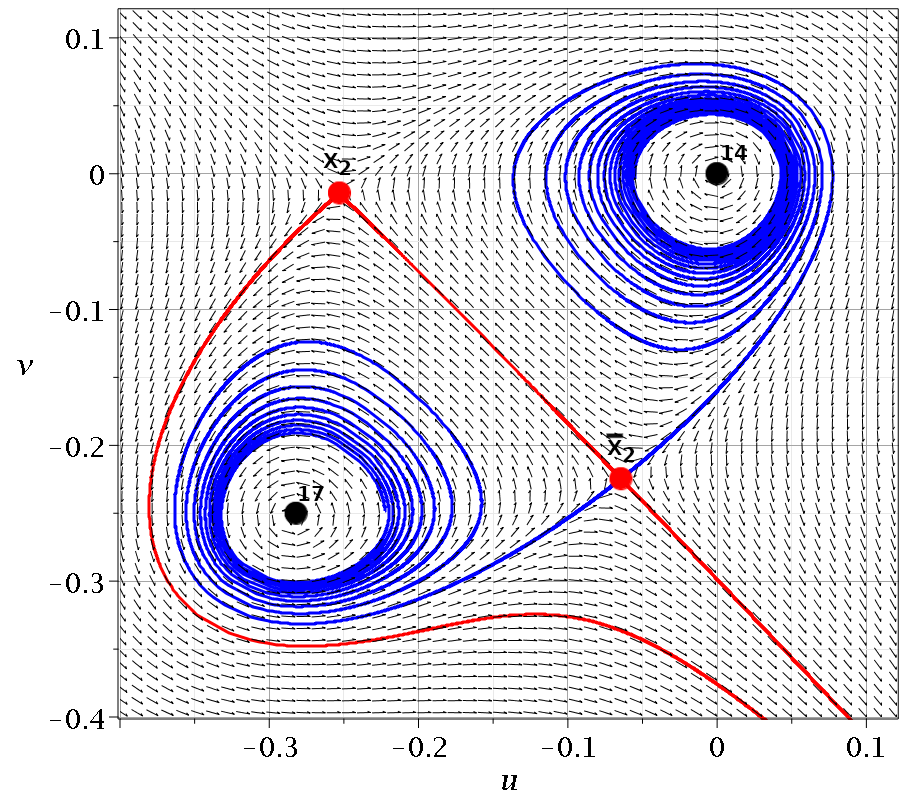}
\caption{Details of the Fig.~\ref{fig10}.}\label{fig11}
\end{figure}

\section{X-points, quantum potential and total potential}

Chaos is introduced when particles approach the X-points close to the nodal points. We will consider in some detail the region close to the nodal points 17 (fixed) and 14 (moving).

In Fig.~\ref{flow} we give the flow around the fixed nodal point, which is at the center ($u=v=0$) of its coordinate system. This frame of reference is identical, in this case, to the inertial frame of reference $(x,y)$, due to the zero velocity of the node. We observe two unstable stationary points, $X_1$ and $\bar{X}_1$. Every X-point has two opposite directions and two opposite unstable directions. Along these directions start asymptotic curves which represent the motion starting very close to X along the stable and unstable directions if we keep the time fixed ($t=0.1$ in this case) and use a fictitious time $s$ that represents an adiabatic approximation \cite{efthymiopoulos2007nodal,efth2009}. Of course as $t$ changes the positions of the X-points and the form of the trajectories of their asymptotic curves change too.

\begin{figure}
\centering
\includegraphics[width=0.7\textwidth]{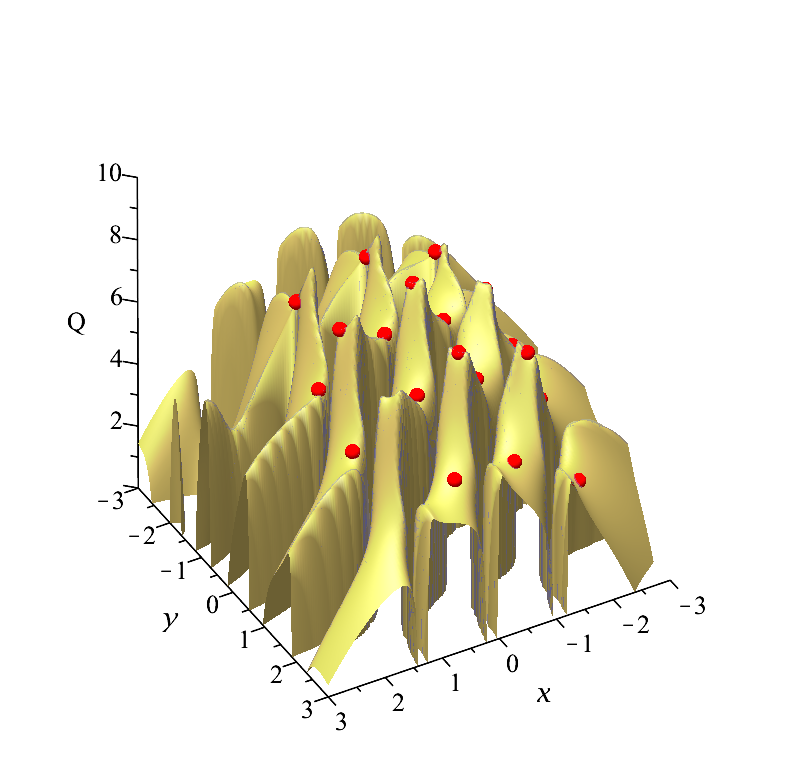}[a]
\includegraphics[width=0.7\textwidth]{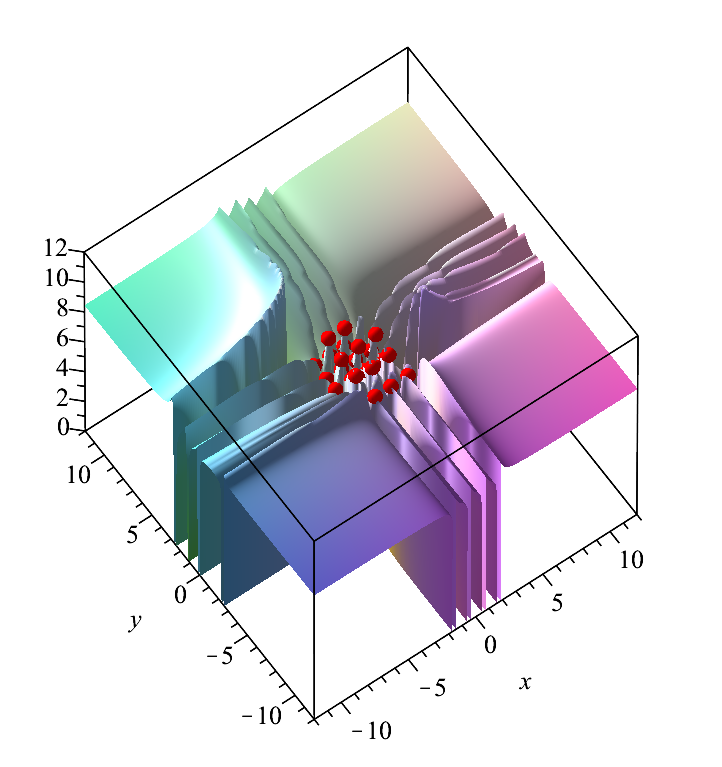}[b]
\caption{a)The quantum potential $Q$ at $t=0.1$ along with several X-points of the wavefunction. We observe that the X-points are very close to the local maxima of $Q$ (red dots) b) The total potential $V_{tot}$, at the same time, has several peaks inside  the central region, but further away  it tends to a small constant value $V_{tot}\simeq 8.4$.
}\label{fig9}
\end{figure}

In Fig.~\ref{flow} the stable asymptotic curves are blue for $X_1$ and green for $\bar{X}_1$. The unstable asymptotic curves are red for $X_1$ and purple for $\bar{X}_1$. Some curves are close to each other and cannot be seen (e.g. the red curve on the left of $X_1$ and the green curve on the right of $\bar{X}_1$). However, if we zoom the regions close to the two X-points, we see that these curves are different.

If now we consider another frame of reference $(u,v)$ around the moving nodal point, we have a figure (Fig.~\ref{fig10}) which is similar to Fig.~\ref{flow} but not identical. In this new frame of reference we find two X-points ($X_2$ and $\bar{X}_2$) which are close but different from $X_1$ and $\bar{X}_1$, and some asymptotic curves seem to overlap. More detailed figures show the small differences. Namely,  the asymptotic curves from ${X}_2$ are shown in Fig.~\ref{fig11}a (stable green and unstable red). The green curve comes very close to $\bar{X}_2$, but does not reach it. The asymptotic curves of $\bar{X}_2$ are shown in Fig.~\ref{fig11}b (stable blue and unstable red). The upper red curve comes very close to ${X}_2$ but does not reach it.

The X-points are remarkable on the surfaces of the quantum potential $Q=-\frac{\hbar^2}{2m}\left(\frac{\nabla^2\Psi}{\Psi}\right)$. The overall form of the surface $Q$ is shown in Fig.~\ref{fig9}a. $Q$ goes to $-\infty$ at the nodal points, but close to them there are various X-points that are close to positive maxima of $Q$. These maxima are abrupt, very close to the infinitely deep tubes around the nodal points. If now we add the classical potential $V$ we have the total potential $V_{tot}=V+Q$ (Fig.~\ref{fig9}b). The X-points are again close to the local maxima of $V_{tot}$. On the other hand,  far from the central region the potential $V$ increases while $Q$ decreases and the total potential $V_{tot}$ tends to a small constant value $V_{tot}\simeq 8.4$. In fact, along most directions the large negative $Q$ and the large positive $V$ almost cancel each other. However, along the $x$ axis there are deep and very thin negative ridges (Fig.~\ref{fig9}b) that extend to infinity, where the large negative value of $Q$ dominates over $V$. Further details about the quantum potential can be found in our recent paper \cite{tzemos2022}

\section{Effectively chaotic trajectories}

If the ratio $\omega_1/\omega_2$ of the basic frequencies is rational all the trajectories are periodic \cite{tzemos2019bohmian}. E.g. if $\omega_y=1$ and $\omega_x=m\omega_y$ then all trajectories are of period $T=2\pi m$. Thus the system is integrable. However it may be difficult or impossible to find an analytic form of the integral. Namely the Bohmian equations of motion give $\dot{x}$ and $\dot{y}$ as functions of $x,y,\cos(t),\sin(t)$, because the functions  $\cos(mt),\sin(mt)$ are expressed as powers of $m$ degree of $\cos(t)$ and $\sin(t)$. These equations give $\cos(t), \sin(t),$ as functions of $x,y,\dot{x},\dot{y}$. Then the expression $\cos^2(t)+\sin^2(t)=1$ is a  function of $x,y, \dot{x}, \dot{y}$. This function is an integral of motion. However in general these functions cannot be given analytically and we have only numerical values of them.

The trajectories in such cases may be very complicated, although they are periodic. Examples of such trajectories have been given in previous papers \cite{contopoulos2008order, tzemos2019bohmian}. 

\begin{figure}
\centering
\includegraphics[width=0.65\textwidth]{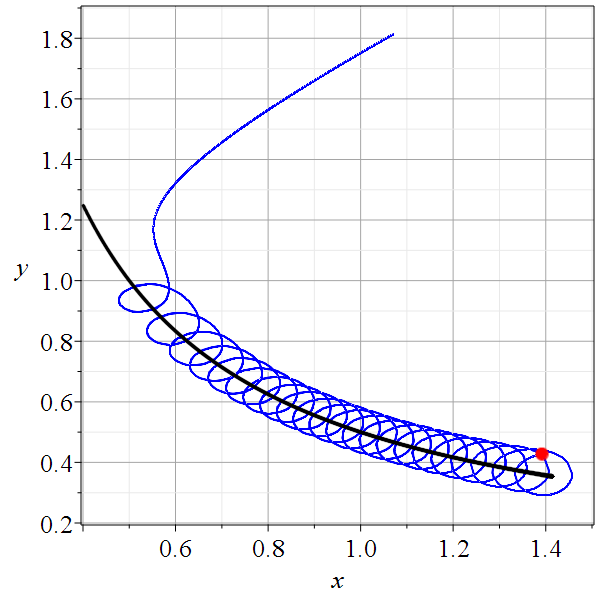}
\caption{A periodic trajectory with $\omega_x=\omega_y=1$ and initial conditions $x(0)=1.0707, y(0)=1.8137$.  The trajectory makes many loops around the nodal point, which moves along a hyperbola (black curve). At $t=T/2=\pi$ (red dot) the trajectory turns back and reaches its initial point at $t=T=2\pi$.}\label{fig13}
\end{figure}

A relatively simple case is shown in Fig.~\ref{fig13}. In this case a periodic trajectory that is trapped for some time around a moving nodal point when $\omega_x=\omega_y=1$. If $\omega_x/\omega_y$ is even slightly different from 1 this trajectory is chaotic. However in the present case the trajectory, after many loops around the nodal point, reaches a limiting point at $t=\pi$ and retraces the same trajectory backwards until it reaches the original point at $t=T=2\pi$. 

Here we can find an analytical form of the integral of motion. The wavefunction is 
\begin{align}
\Psi=\Psi_{0,0}+\Psi_{1,0}+\Psi_{1,1},
\end{align}
Thus the Bohmian equations of motion are
\begin{eqnarray}
&\dot{x}=\frac{\sqrt{2}\sin(t)+2y\sin(2t)}{G}\\&
\dot{y}=\frac{2x\left(\sqrt{2}x\sin(t)+\sin(2t)\right)}{G},
\end{eqnarray}
where
\begin{equation}
G=1+2x^2(1+2y^2)+2\sqrt{2}x(1+2xy)\cos(t)+4xy\cos(2t).
\end{equation}
On the other hand the nodal point is at
\begin{equation}
x_N=-\frac{\sin(2t)}{\sqrt{2}\sin(t)}, \quad y_N=-\frac{\sin(t)}{\sqrt{2}\sin(2t)}
\end{equation}
Thus the nodal point describes the hyperbola
\begin{align}
x_Ny_N=\frac{1}{2}
\end{align}
From the equations (24) and (25) we find
\begin{equation}
\frac{\dot{x}}{\dot{y}}=\frac{1+2\sqrt{2}y\cos(t)}{\sqrt{2}x\left(\sqrt{2}x+2\cos(t)\right)}
\end{equation}
and from this equation we derive $\cos(t)$ as a function of $x,y,\dot{x},\dot{y}$
\begin{align}
\cos(t)=\frac{2x^2\dot{x}-\dot{y}}{2\sqrt{2}(y\dot{y}-x\dot{x})}
\end{align}
Introducing this value in Eq.(24) we find
\begin{equation}
\sin(t)=\frac{\dot{x}G}{\sqrt{2}(1+2\sqrt{2}y\cos(t))}=\frac{G(y\dot{y}-x\dot{x})}{\sqrt{2}(2x^2y-x)}
\end{equation}
Thus the integral of motion is $\sin^2(t)+\cos^2(t)=1$.
This is a function of $x,y, \dot{x},\dot{y},t$  is  rather complicated but it has been checked numerically.

However in more general cases of rational frequencies the functions are even more complicated and may not be able to be written analytically. E.g. in the case of two qubits \cite{tzemos2019bohmian}, although the case $\omega_x=\omega_y=1$ is very simple and gives $x-y=const$, for more complex values $\omega_x/\omega_y=rational$ we cannot find analytical forms for $\cos(t)$ and $\sin(t)$. This system is integrable but we cannot give the analytical forms of the integral.

\section{Conclusions}

We considered the Bohmian trajectories of a 2-d quantum harmonic oscillator with non commensurable frequencies, guided by wavefunctions of the form $\Psi=a\Psi_{m_1,n_1}+b\Psi_{m_2,n_2}+c\Psi_{m_3,n_3}$ where $\Psi_{m,n}=\Psi_m(x)\Psi_n(y)$, with $m, n$  relatively large quantum numbers which lead to the existence of multiple nodal points.

\begin{enumerate}
\item We  find first the trajectories of the nodal points where $\Psi=0$. Analytic solutions can be found for relatively small $m$ and $n$. However, an important class of solutions is found if two $m's$ or $n's$ are equal. In such a case the nodal points are of two types:  (a) Fixed nodal points in the coordinate system $(x,y)$ (time independent) and b) moving nodal points.
\item In the case where $m_1=m_2$, the number of fixed points depends on the $m_1$ (the order of the equal degrees) and on the degree $n_3$ of the term $\Psi_{m_3,n_3}$. We have $n_3$ sets of $m_1$ fixed points. Thus the total number of fixed points is $m_1n_3$. In particular, if $m_1=0$ or $n_3=0$, we have no fixed points at all. 

\item At particular times  the moving nodal points go to infinity. Furthermore at special times some moving points collide with particular fixed points and then they continue moving. This type of collisions is different from a previously considered collision form between two moving nodal points which leads to their disappearance. Such a case was considered in \cite{efth2009}.

\item  In a particular example we have  $m_1=m_2=3$ and $n_3=5$, i.e. a total number of 15 fixed nodal points.  In this example we  have four quadruplets of moving nodal points. Thus the total number of nodes is $31$.

%

\item  Then we studied the trajectories of quantum particles. In general when we have many nodal points, the trajectories are chaotic. Chaos is introduced close to the nodal points. In fact near a nodal point there is one or more unstable points, stationary in the frame  of the nodal point, called X-points, that have two stable and two  unstable directions. Any particle approaching an X-point close to its stable direction is deflected along one or the other of the unstable directions. This mechanism is the same as the one considered in the case of a single nodal point \cite{efthymiopoulos2007nodal} and an infinity of nodal points \cite{tzemos2019bohmian,tzemos2020ergodicity}.

\item The trajectories that approach a nodal point form spiral loops around it for some time but then they go far from the nodal point. We study their phenomenology in detail. Of special interest are the trajectories that approach a couple of nodal points when they are close to each other.

\item  We calculated the quantum potential $Q$ and the total potential $V_{tot}=Q+V$. $Q$ tends to $-\infty$ at  the nodal points, while the nearby X-points are always close to local positive maxima of $Q$ and $V_{tot}$.

\item When the ratio of the basic frequencies is a rational number all the trajectories are periodic. However many trajectories are ``effectively chaotic'', i.e. they behave as chaotic for a long time.

\end{enumerate}

\section*{Acknowledgments}
This research was conducted in the framework of the program of the RCAAM of the Academy of Athens ``Study of the dynamical evolution of the entanglement and coherence in quantum systems''.

\section{Appendix: Finding the nodal points}\label{secA1}

The nodal points are of fundamental importance in the study of Bohmian chaos. These points  are  mathematical singularities of the Bohmian flow. Quantum particles close to a nodal point have large velocities, forming  spiral vortices around it for some time. However, later the particles escape from the neighbourhood of the node. These esapes happen in two cases
\begin{enumerate}
\item when the nodal point acquires a large velocity when going or coming from infinity
\item when a moving nodal point approaches and collides with a fixed (non moving) nodal point. This second mechanism appeared for the first time in the present paper.
\end{enumerate}
After the escape the particle wanders around the configuration space until it comes close to the same or another nodal point and so on. 

The evolution of the nodal points is not dictated by the Bohmian equations of motion but from their defining set of equations 
\begin{equation}
\Psi_{Real}(x, y,t)=\Psi_{Imaginary}(x, y,t)=0.\label{de}
\end{equation}

The solutions of these equation become singular from time to time. These are the times where the nodal points go to or come from infinity. 


However, the solutions of these equations can be found analytically only if the quantum numbers $m$ and $n$ are small. But if $m$ and $n$ are large, then in most cases the nodal points are found only numerically. Namely, one needs  first  to detect graphically where the velocities of the Bohmian flow form vortices (as in  Fig.~\ref{fig2})  and then to plot successively the Bohmian velocity  field by gradually increasing the time  and spot the moving nodal points at the centers of the vortices. 

In the present paper we followed this method and put  the successive figures into video simulations, where the trajectories of the nodal points   became evident. Furthermore, in the same way we found the positions of the X-points by observing the points from which emanate the  two stable and two unstable eigendirections.

\bibliographystyle{elsarticle-num}
\bibliography{sn2}


\end{document}